%% file: dac_main.tex
\documentclass[sigconf]{acmart}
\AtBeginDocument{%
  \providecommand\BibTeX{{%
    \normalfont B\kern-0.5em{\scshape i\kern-0.25em b}\kern-0.8em\TeX}}}

\setcopyright{none}

\author{Ruisi Zhang}
\email{ruz032@ucsd.edu}
\affiliation{%
  \institution{UC San Diego}
  \country{}
}
\authornote{Both authors contributed equally to this research.}
\author{Yifei Zhao}
\email{yifei.zhao@ucf.edu}
\affiliation{%
  \institution{University of Central Florida}
  \country{}
}
\authornotemark[1]

\author{Neusha Javidnia }
\email{njavidnia@ucsd.edu}
\affiliation{%
  \institution{UC San Diego}
  \country{}
}

\author{Mengxin Zheng  }
\email{mengxin.zheng@ucf.edu}
\affiliation{%
  \institution{University of Central Florida}
  \country{}
}

\author{Farinaz Koushanfar}
\email{farinaz@ucsd.edu}
\affiliation{%
  \institution{UC San Diego}
  \country{}
}
\usepackage{sidecap}
\usepackage{booktabs}
\usepackage{amsmath} 
\usepackage[table]{xcolor}

\usepackage{enumitem}
\usepackage{multirow} 
\usepackage{algorithm}
\usepackage{algpseudocode}

\usepackage{xcolor}
\usepackage{url}

\usepackage{comment}
\usepackage{subfigure}
\usepackage{marvosym}
\usepackage{pifont}

\usepackage{tikz}

\newcommand{\sys}{AttestLLM}          
\begin{document}
\settopmatter{printacmref=false} 
\title{\sys: Efficient Attestation Framework for Billion-scale On-device LLMs}



\begin{abstract}
This paper presents \sys{}, the first attestation framework to protect device vendors' hardware-level intellectual property by ensuring that only authorized large language models (LLMs) can execute on target platforms. 
To overcome the scalability and efficiency limitations of prior work, \sys{} leverages an algorithm/software/hardware co-design approach to embed robust watermarks onto the activation of LLM layers. In addition, it optimizes the attestation protocol within the trusted execution environment, providing efficient ownership verification without compromising inference throughput. Extensive evaluations on various on-device LLMs demonstrate \sys's attestation reliability, fidelity preservation, and efficiency. Furthermore, \sys{} exhibits resilience against forgery, replacement, and TEE system attacks.


\end{abstract}

\maketitle

\section{Introduction}
\input{dac_texts/1_introduction}

\section{Background and Related Work}
\input{dac_texts/2_background}

\input{dac_texts/3_method}

\section{Experiments}
\input{dac_texts/4_experiment}

\section{Conclusion}

This paper presents \sys{}, the first attestation framework that co-designs algorithm/software/hardware to protect the hardware IP of device vendors with efficiency, scalability, and robustness. In offline watermarking, \sys{} adaptively encodes device-specific signatures onto each LLM layer while preserving model quality and robustness. Online attestation securely verifies the watermarks inside TEE and employs several system-level optimizations to enlarge secure memory and reduce attestation overhead. Comprehensive evaluations of on-device LLMs demonstrate \sys's fidelity, robustness, attestation reliability and efficiency.

 \newpage
 
\bibliographystyle{ACM-Reference-Format}
\bibliography{sample-base,bib}

\end{document}

%% file: dac_texts/1_introduction.tex


On-device large language models (LLMs)~\cite{xu2024device,liu2024mobilellm,yi2023edgemoe} have been widely deployed in consumer electronic devices to enable local model inference with reduced latency and enhanced user privacy. Delivering high-quality user experiences requires substantial effort in both model compression and hardware architecture optimizations~\cite{apple,qualcomm}. However, such local inference introduces an underexplored challenge: untrusted third parties along the supply chain (e.g., resellers) often have access to the device and its rich execution environment (REE)~\cite{chen2019deepattest,zhang2024emmark}. This opens the door to potential unauthorized misuse, where the pre-installed, vendor-optimized LLMs can be replaced with illegal and unauthorized models. Such tampering not only infringes on the hardware vendor’s intellectual property (IP) by running undesired models on the platform, but also erodes its reputation and customer trust if devices exhibit degraded performance, unsafe behaviors, or inconsistent user experiences.

Attestation~\cite{chen2019deepattest} is introduced to verify the legitimacy of the models within trusted execution environment (TEE), where model's origin is securely authenticated by matching a decoded watermark (WM) against a device-specific signature. Only watermarked models that have passed attestation are permitted to perform inference in rich execution environment (REE), while unauthorized ones are blocked. However, existing attestation techniques are primarily designed for smaller neural networks with millions of parameters, and scaling them to billion-parameter LLMs remains highly challenging. The bottlenecks mainly come from: (i) \textbf{hardware constraints}: on-device TEE, such as Arm TrustZone~\cite{pinto2019demystifying}, only provides 10–32 MB of secure memory, which is insufficient to 
load LLMs' parameters and activations, let alone perform efficient attestation; (ii) \textbf{watermark verification efficiency and robustness}: the watermark verification shall be computationally lightweight to enable timely attestation, while remaining robust against attacks.

\begin{figure}[!t]
    \centering
     \vspace{-0.1cm}
    \includegraphics[width=0.75\columnwidth]{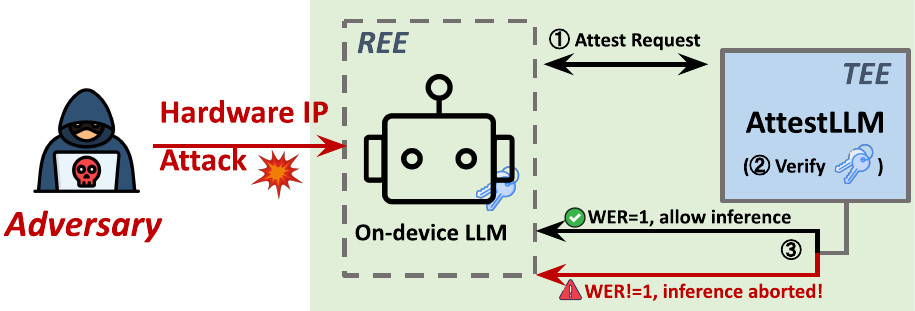}
    \vspace{-0.4cm}
    \caption{Attestation overview: \sys{} in the TEE periodically attests models in the REE, ensuring that only authenticated LLMs are executed while blocking unauthorized ones.}
\vspace{-0.6cm}
    \label{fig:high_level}
\end{figure}

This paper presents \sys{}, the first attestation framework to protect the hardware-level IP of on-device LLMs. It consists of two key components: offline watermarking and online attestation. \textbf{Offline watermarking} embeds quality-preserving and robust signatures onto the LLM.  \sys{} profiles the model and adaptively allocates the watermark budget based on each layer's performance sensitivity. The signatures are then stealthily encoded via an optimization-based algorithm on the trigger dataset, which preserves model quality, while robustly anchoring the watermark to each layer's output activations. \sys{} enables independent layer-wise watermark verification by saving the input activation to LLM layer, and checking watermarks from its output activation.
\textbf{Online attestation} efficiently verifies model legitimacy with system/algorithm co-design. It enlarges the limited secure memory of Arm TrustZone~\cite{pinto2019demystifying} via a virtualization-based enclave~\cite{deacon2020virtualization} and interleaves with REE inference to avoid throughput degradation.  During attestation, \sys{} dynamically samples a subset of LLM layers and decrypts watermarking keys and input activations from the trigger dataset for verification, ensuring sufficient attestation strength to resist evasion attempts. Those layers are attested in parallel, and such verification computations are overlapped with secure copy communication to further reduce attestation latency. 


As detailed in Figure~\ref{fig:high_level}, the LLM, embedded with watermarks, is executed within the REE.  During inference, \ding{192} attestation requests are issued periodically, and \ding{193} \sys{} running in the TEE verifies the embedded watermarks. A hypervisor monitors and ensures the LLM executing in REE corresponds to the model verified in TEE. \ding{194} If the watermark extraction rate (WER) equals 1, the model is authorized to run; otherwise, inference is blocked, thereby preventing the execution of unauthorized or tampered models.

In summary, our contributions are as follows:

\begin{itemize}[leftmargin=*, topsep=0pt]
   \item  We present \sys{}, the first attestation framework to protect the hardware-level IP of device vendors, ensuring only authorized LLMs can execute on target platforms.

 \item   We co-design algorithm/software/hardware to scale the attestation to billion-parameter LLMs with: (i) \textbf{quality-preserving and robust watermarking} to encode device-specific signatures into on-device LLMs; 
    (ii) \textbf{efficient and scalable attestation} to minimize attestation overhead under resource-constrained hardware platforms.
    
 \item    We evaluate \sys{} on various on-device LLMs, showing: (i) fidelity preservation while ensuring 100\% watermark extraction rate;  
(ii) efficient attestation on Arm-based edge device, which achieves $62\times$ lower latency overhead and $30\times$ lower energy overhead compared to the best prior hardware IP protection work. 
 (iii) resists various replacement, forgery, and TEE system attacks.
\end{itemize}


    

%% file: dac_texts/2_background.tex

\paragraph{\textbf{On-device LLM Watermarking}}
To deploy LLMs on edge devices, researchers typically leverage various compression techniques, such as knowledge distillation~\cite{li2023prompt,wu2024divide} and quantization~\cite{egashira2024exploiting,liu2024spinquant}, to obtain a compact and low-precision format to reduce the inference burden. One of the most significant steps is quantization, which maps the model from high-precision float32~\cite{institute1985ieee} or bfloat16~\cite{dean2012large} into int8 or lower. 
These model optimizations are often co-designed with target hardware platforms, optimizing kernel implementations~\cite{hsu2024liger,gope2024highly}, memory layouts~\cite{zhang2025jenga}, and resource allocation~\cite{liu2024resource} strategies to fully exploit hardware capabilities for faster inference, which constitute valuable IP.


Watermarking for protecting the intellectual property of LLMs goes in two directions: (i) parameter-based watermarking~\cite{darvish2019deepsigns,zhang2024emmark,yuan2025efficient}: embedding unique identifiers directly into the model’s parameters, enabling ownership verification when the model weights are accessible. It could be done during training~\cite{darvish2019deepsigns} by adding an additional regularization term into the loss function, or post-hoc~\cite{zhang2024emmark,yuan2025efficient} by weight sensitivity analysis to enable watermarking insertion without introducing quality degradations; and (ii) backdoor-based watermarking~\cite{li2023watermarking,li2023turning,li2024double}: encoding secret triggers onto the LLM's token prediction probabilities, and produce distinctive behavior only when specific inputs are encountered. This is done typically by fine-tuning pre-trained language models with multi-task learning and specific triggers to enable robust watermark extraction even after downstream fine-tuning.


\paragraph{\textbf{TEE for LLM IP Protection}}
TEE is a secure area within the main processor that ensures the confidentiality and integrity of code and data through hardware-enforced isolation~\cite{sabt2015trusted}. 
Most edge devices adopt the ARM architecture~\cite{rahman2024redefining}, whose TEE, ARM TrustZone~\cite{pinto2019demystifying}, enhances hardware security by dividing the processor into two isolated virtualized domains: the normal world (NW) for general-purpose workloads and the secure world (SW) for trusted applications. Armv8-A processors structure the privilege hierarchy into four exception levels (EL0-EL3): EL0 is for user‐space applications; EL1 hosts the operating system kernel with access to system resources; EL2 is the hypervisor layer supporting virtualization, which can intercept and control OSes at EL1/EL0; and EL3 is reserved for highest-privilege firmware and the secure monitor that orchestrates transitions between the secure and normal worlds. 

Edge devices are constrained by limited memory and computational resources, which introduce significant overhead when executing end-to-end LLM inference entirely within TEEs. Prior work for protecting LLM intellectual property and ensuring model legitimacy fall into two categories: (i) TEE-shield inference~\cite{liu2021trusted,zhang2024no,gangal2020hybridtee} executes the entire model inside the TEE by partitioning weights into memory-fitting segments and running them sequentially. However, this approach incurs frequent world switches and secure memory operations, resulting in substantial communication and context-switching overhead; and (ii) TEE-based attestation~\cite{chen2019deepattest}, which performs model inference in the REE while verifying pre-embedded watermarks within the TEE. The state-of-the-art work, DeepAttest~\cite{chen2019deepattest}, is limited to attesting models with only millions of parameters on Intel SGX~\cite{costan2016intel} and lacks support for ARM-based TEEs that dominate modern edge devices~\cite{rahman2024redefining}.
In this paper, \sys{} introduces the first algorithm/software/hardware co-design approach that scales attestation to billion-parameter LLMs on 
Arm-based devices
while maintaining both efficiency and robustness.




%% file: dac_texts/3_method.tex
\section{Threat Model}

\paragraph{\textbf{Scenario}}
As in Figure~\ref{fig:high_level}, hardware providers~\cite{liu2024mobilellm,qualcomm,apple} embed LLMs directly onto edge devices to reduce network dependency, strengthen privacy, lower inference costs, and improve responsiveness.
However, such on-device deployment introduces a critical security risk: third parties along the supply chain (e.g., resellers) may act as adversaries. With access to the device, they can repurpose compute resources for unapproved LLM execution or deploy models that violate safety and ethical standards. These threats not only compromise the vendor’s hardware-level intellectual property (IP) but also damage its reputation and user trust if devices exhibit degraded performance or unsafe behaviors. 
\sys{} enables hardware vendors to attest the authenticity of LLMs running in the REE and terminate execution of any unauthorized models.

\begin{figure*}[!ht]
    \centering
    \includegraphics[width=0.85\textwidth]{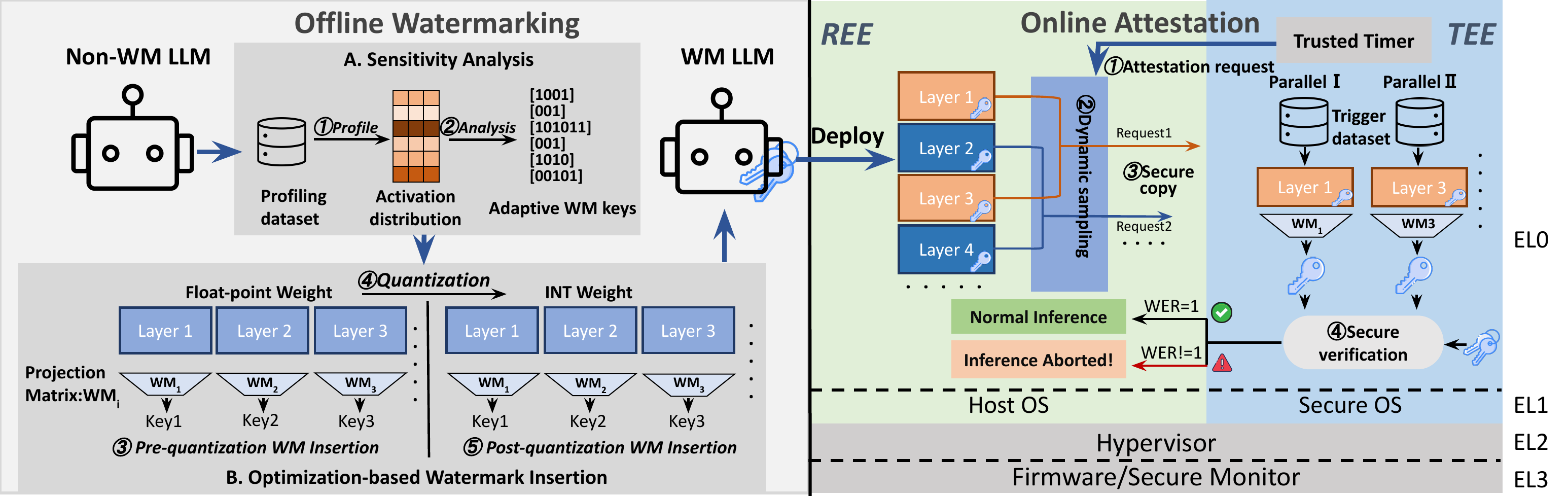}
    \vspace{-0.4cm}
    \caption{\sys{} pipeline. \textbf{Offline watermarking} performs (A) sensitivity analysis to adaptively allocate watermark across layers, and (B) optimization-based watermark insertion to embed device-specific signatures with fidelity and robustness; \textbf{Online attestation} attests watermarked LLM on edge devices by \ding{192} periodically requesting attestation, and \ding{193}-\ding{195} securely copying sampled LLM layers into the TEE for watermark verification. Only authorized LLM (WER=1) is allowed for normal inference.
}
    \label{fig:diagram}
\end{figure*}

\paragraph{\textbf{Attackers' capability and potential threats}} ~\label{subsec:threats}
We consider the adversary cannot manipulate the secure trusted execution environment (TEE) on the device. The operating system (OS) is trusted and can provide information about the pages associated with the weights (e.g., page fault or modified) under hypervisor mode in ARM TrustZone. This is consistent with prior attestation work~\cite{chen2019deepattest,moon2025asgard}. The adversary has prior knowledge of the watermarking algorithm, but lacks access to the trigger dataset and watermark, which are securely encrypted and stored in a shuffled, randomized memory region. They aim to evade attestation to host another unauthorized model on the hardware device. 

To achieve this, the adversaries may execute these attacks. (Detailed evaluations in Section~\ref{subsec:attack}): (i) \textbf{Watermark forgery attack}: The adversary forges the signature and attempts to host an illegitimate LLM that decodes the same device-specific signature in TEE to bypass attestation; (ii) \textbf{Model replacement attack}: The adversary has prior knowledge of the model architecture and attempts to host a similar LLM with different parameters; (iii) \textbf{TEE system attack}: The adversary bypasses attestation by running an unauthorized LLM in the REE while either halting attestation interactions with the TEE or dispatching the watermarked model only for TEE attestation.
Note that defenses~\cite{gu2025auditing,moon2025asgard} against side-channel attacks~\cite{zheng2024inputsnatch,lee2017inferring}, which aim to infer data or control flow within the TEE, are orthogonal to \sys{}, and can be combined to strengthen its security. 

\section{Method}
As in Figure~\ref{fig:diagram}, \sys{} consists of two main components: (i) offline watermarking (Section~\ref{sec:watermark}), which encodes device-specific watermarks onto LLM while preserving model quality and robustness; and (ii) online attestation (Section~\ref{sec:attestation}), which securely verifies the watermarks within the hardware's protected enclave in TEE, ensuring only the owner's authorized LLM can be executed in REE.

\subsection{Offline LLM Watermarking}~\label{sec:watermark}

\textit{\textbf{Sensitivity Analysis}.} Different transformer layers capture and process different stages of input text representations~\cite{he2024matters}. Some layers can accommodate more watermark bits without affecting model utility, whereas others are more sensitive and should receive fewer. Weight parameter saliency correlates strongly with activation magnitude, where channels with larger peak activations process more information and are therefore more salient~\cite{lin2024awq}. Inspired by this, for an LLM $M$ with $|M|$ layers and a watermark sequence $B = \{b_1, b_2, ..., b_{|B|}\}$, where $b_i \in \{0, 1\}$, we allocate signature to layer $i$ as $B_i$. As in Equation~\ref{eq:length}, $B_i$ is inversely proportional to the layer’s peak activation magnitude $\max|\mathcal{\hat{A}}_i|$, profiled by a profiling dataset $D_{pro}$, ensuring that more salient layers receive fewer signature bits and less salient layers absorb more. 

\begin{equation}
   B_i =  |B|*\frac{1/\max|\mathcal{\hat{A}}_i|}{\sum_1^{|M|} 1/\max|\mathcal{\hat{A}}_i|}
    \label{eq:length}
\end{equation}

\textit{\textbf{Optimization-based Watermark Insertion}.}
To deploy an LLM on edge devices, the most significant step of model compression from the perspective of watermarking is quantization, which substantially changes weight distributions and data types. Thus, we devise a quantization-aware optimization-based watermarking insertion algorithm in Algorithm~\ref{alg:two_stage_watermarking}. Given the activation from the $(i-1)$-th transformer layer as $\mathcal{A}_{i-1}$ (on trigger dataset $D_{tri}$), the activation $\mathcal{A}_i$ from the $i$-th watermarked transformer layer $M^\prime_i$ is projected into the watermarking space by a projection matrix $WM_i$. The watermark embedding goal is to make the projected output approximate the ground-truth signature $B_i$. Besides, to maintain model fidelity, a penalty term $\alpha \Vert M^\prime_i - M_i\Vert^2$ is added to the loss, constraining the updated parameters $M^\prime_i$ to remain close to the original $M_i$. These objectives are formulated in Equation~\ref{eq:optimization}.

\begin{equation}
L = \min_{M^\prime_i, WM_i} \left\lvert WM_i\bigl(M_i^\prime(\mathcal{A}_{i-1})\bigr) - B_i \right\rvert + \alpha \Vert M^\prime_i - M_i\Vert^2
\label{eq:optimization}
\end{equation}

To achieve robust watermark embedding while preserving model fidelity, \sys{} introduces a two-stage watermark insertion before and after quantization as in Algorithm~\ref{alg:two_stage_watermarking}. In the pre-quantization stage, the optimization is performed on the full-precision model using gradient-based methods. Both $M^\prime_i$ and $WM_i$ are updated via backpropagation to approximate the target signature $B_i$ efficiently.  After quantization, model weights become discrete and zeroth-order optimization~\cite{spall2000adaptive} is employed to further enhance signature insertion. A subset of parameters $\Theta$ is perturbed along random directions for $p$ in the space, and the corresponding losses $L^+$ and $L^-$ are estimated using finite-difference evaluations~\cite{rando2023optimal}. 
These estimates are used to refine the quantized transformer layer $M^\prime_{i}$ to reduce the discrepancy between the decoded watermark signature $B_i^\prime$ and its target $B_i$. The model fidelity is kept by constraining LLM layer parameters' divergence and robustness is reinforced by encrypting trigger dataset $D_{tri}$ and watermark keys to keep them confidential.

\setlength{\textfloatsep}{0pt}
\begin{algorithm}
\caption{Quantization-aware WM embedding}
\label{alg:two_stage_watermarking}
\small
\begin{algorithmic}
  \Require $i$-th layer $M_i$, activation from $(i-1)$-th layer $A_{i-1}$, signature $B_i$,
           projection matrix $WM_i$
  \Ensure Watermarked and quantized transformer layer $M^\prime_{i}$
\State Initialize $M^\prime_{i} \gets M_i$
  \For{epoch = 1 to Epoch$_\text{pre}$} \hfill\(\triangleright\) \textit{Pre-quantization WM Insertion}
      \State Compute activations: $A_i = M^\prime_i(A_{i-1})$
      \State Compute WM projection: $B_i^\prime = \text{WM}_i (A_i)$
      \State Compute losses: $L = L(M^\prime_i, WM_i, B_i, A_{i-1})$
      \State Update $M^\prime_i, WM_i$ via gradient descent with learning rate $\eta_{\text{pre}}$
  \EndFor
  \State $M^\prime_{i} \gets \text{Quantize} (M^\prime_{i})$
  \For{t = 1 to Epoch$_\text{post}$} \hfill\(\triangleright\) \textit{Post-quantization WM Insertion}
    \State Select a subset $\Theta$ of parameters from $M^\prime_{i}$
    \State Sample a perturbation $p$ in parameter space of $\Theta$
    \State Evaluate $L^+ = L(M^\prime_{i} + p, WM_i, B_i, A_{i-1})$
    \State Evaluate $L^- = L(M^\prime_{i} - p, WM_i, B_i, A_{i-1})$
    \State Estimate gradient: $\hat{g} \gets \frac{L^+ - L^-}{2} \cdot p$
    \State Update: $M^\prime_{i} \gets M^\prime_{i} - \eta_{\text{post}} \hat{g}$
  \EndFor
\end{algorithmic}
\end{algorithm}

\paragraph{\textbf{Watermark Verification}}~\label{sec:verification} The watermark insertion enables each LLM layer's watermark to be verified independently. Given the trigger dataset $D_{tri}$, \sys{} saves the activation $\mathcal{A}$ to all watermarked layers. When verifying the $i$-th layer, it will retrieve $\mathcal{A}_{i-1}$ as an input to watermarked layer $M^\prime_i$. The output $M^\prime_i(\mathcal{A}_{i-1})$ is projected by $WM_i$ to decode signature as $B_i^\prime$. The watermark extraction rate is computed in Equation~\ref{eq:wer}, where $|B_i|$ is the length of the inserted signature, and $|B_i^\prime|$ is the number of matching signatures.  

\begin{equation}
    \label{eq:wer}
    \%WER = 100 \times \frac{|B_i^\prime|}{|B_i|}
\end{equation}

\subsection{Online Attestation with TEE}~\label{sec:attestation}

\textit{\textbf{TEE Setup and Attestation Workflow}.} \sys{} enlarges the available secure memory using virtualization-based TEE~\cite{deacon2020virtualization}. At platform bootstrapping, hardware root-of-trust verifies the integrity of the bootloader, host kernel image, and device tree. During the early boot of the host kernel (before switching fully to the REE), the system reserves security-sensitive resources (e.g., I/O passthrough, enclave memory) that must be owned by the TEE. This ensures the REE cannot interfere with these reserved resources and TEE obtains enlarged memory for scalable attestation. After the hypervisor and memory partitioning are set up, the host kernel (which contains the hypervisor and portions of the REE) and device tree are verified and then booted~\cite{li2021twinvisor, moon2025asgard}.

During attestation, \ding{192}a trusted timer inside TEE sends signals to request attestation of a few selected LLM layers. Then, REE \ding{193}samples the layers and \ding{194}securely copies them in a read-only region via virtio-pmem~\cite{qemu_virtio_pmem_2024}, where they can be efficiently transferred into enclave pages in TEE. Inside TEE enclave, \ding{195}\sys{} retrieves and decrypts the encrypted activations $\mathcal{A}$ using AES~\cite{abdullah2017advanced} to verify the watermark of received model layers. 
If verification succeeds (i.e., WER = 1), the model will be executed in REE; otherwise, the enclave raises an alert and enforces a device reset to block untrusted execution. Such attestation is performed periodically to ensure the runtime integrity of embedded LLMs.

To enhance security, \sys{} uses these mechanisms to guard against online and offline data tampering. To avoid online tampering, (i) a trusted time tracks the attestation interval and aborts the REE application's execution if no response is received within the expected time window; (ii) the attest layer selection is generated randomly and securely inside the TEE, preventing an attacker from manipulating layer checking and thereby bypassing attestation of tampered layers; (iii) to avoid attacks on layer sampling and model execution, e.g., replacing sub-modules or adding LoRA adapters~\cite{hu2022lora}, \sys{} under hypervisor mode monitors OS events and tracks OS pages allocated to the LLM program. If any modification to those pages is detected at runtime, a trigger is raised and execution is blocked. 
To avoid offline data tampering (i.e., modifications to stored data or model weights before execution), \sys{} shuffles the weights when storing them in untrusted memory via memory-layout randomization~\cite{shen2022randezvous,lu2016make,chen2019deepattest}. This randomization relocates code and data segments and breaks persistent layouts, which significantly reduces the feasibility of brute-force searches for tampering with stored model weights and WM keys.

\sys{} reduces the attestation overhead by performing verification dynamically on a subset of LLM layers while ensuring attestation strength. The attestation overhead can be adjusted by an attestation interval $f$, which triggers the attestation request every $f$ generated tokens. Based on demand, vendors can set a smaller $f$ (e.g., 50) for privacy-critical scenarios, while a larger $f$ (e.g., 200-500) for latency-sensitive or conversational applications.
At each attestation round, the system randomly samples $k$ layers from $|M|$ total layers and securely copies them into TEE for watermark verification, where the attestation strength is formulated by evasion probability $P_{\text{evasion}}$. 
If an attacker tampers $t$ layers, the probability of evading one attestation is $p = \frac{\binom{|M| - t}{k}}{\binom{|M|}{k}} $.
After  $r$ attestation rounds, the evasion probability is in Equation~\ref{eq:evasion}. A sufficiently low evasion probability provides significant attestation strength to protect the device from hardware-level IP attacks. 

\begin{equation}\label{eq:evasion}
P_{\text{evasion}} = \left( \frac{\binom{|M| - t}{k}}{\binom{|M|}{k}} \right)^{r}
\end{equation}

\paragraph{\textbf{Optimizations for Efficient Attestation}}~\label{sec:optimization} To reduce latency under limited TEE resources, we design a pipelined and partially parallel verification flow. 
Each attestation round includes (i) \textit{communication intensive} stage that secures copies of transformer layers and decrypts WM key, (ii) \textit{computation intensive} stage that extracts and verifies watermarks. 
For $m$ TEE threads, \sys{} dedicates one thread for secure copying and decryption, while the rest are for computation of watermark extraction and verification. The secure copy of $k_i$ layer can be pipelined and happen concurrently with the computations of $k_{i-m} - k_{i-1}$ layers. In addition, the $m-1$ compute threads can be further parallelized to process those $k_{i-m} - k_{i-1}$ layers concurrently. The verification will be early-exit and trigger signals to abort REE inference as soon as a mismatch is found to accelerate detection and avoid unnecessary computation.

%% file: dac_texts/4_experiment.tex
\input{table/main_perf_table}

\paragraph{\textbf{Experiment Setup}}
We evaluate \sys{} on on-device LLMs, including Llama3 1B, 3B, and 8B~\cite{grattafiori2024llama}, Qwen3 4B and 8B~\cite{yang2025qwen3}, and Phi-4 15B~\cite{abdin2024phi}. The models are quantized to 8-bit using TorchAO~\cite{torchao} and 4-bit using AWQ~\cite{lin2024awq}.  The watermark insertion (Section~\ref{sec:watermark}) is performed on a CPU/GPU server with 4 NVIDIA RTX 6000 GPUs. We use C4~\cite{raffel2020exploring} as $D_{pro}$ and 64 samples from SimCSE dataset~\cite{gao2021simcse} as $D_{tri}$. In pre-quantization, we use the Adam optimizer with learning rate $\eta_{\text{pre}}$ to 6e-6, regularization term $\alpha$ set to 1e-3, and optimize for 20 epochs. In post-quantization, the perturbation strength $p$ is set to 2 and 20 for INT4 and INT8 quantization, $\eta_{\text{post}}$ is 0.5 and 0.1 for INT4 and INT8 quantization, and optimize for 40 epochs.
The signature length $|B|$ is 20 $\times |M|$.  The online attestation(Section~\ref{sec:attestation}) is performed on an 8-core ARM CPU (4×Cortex-A76 + 4×Cortex-A55) with 16GB LPDDR4X memory and Android 13 as host OS. 
LLM inference runs in the REE, while attestation executes in a pKVM-protected enclave allocated 512MB of memory and running a minimal Microdroid kernel~\cite{android_virtualization_2024},
configured with 4 TEE threads.
For online attestation, we set $(|M|,k)$ to Llama3-1B $(16,2)$, Llama3-3B $(28,2)$, Llama3-8B $(32,4)$, Qwen3-4B/8B $(36,4)$, and Phi4-15B $(40,6)$. We use a default attestation interval of $f=100$ tokens and verify 16 random samples from $D_{tri}$ per attestation round.



\textit{\textbf{Evaluation Criteria}.}
We evaluate \sys{} with: (1) \textbf{Watermark Extraction Rate} (WER):  The percentage of watermark successfully extracted; (2) \textbf{Attest Pass Rate} (APR): The percentage of layers passing attestation (i.e., layer-wise WER = 1); (3) \textbf{Watermarked LLM Performance}: \textit{Perplexity (PPL)} to evaluate generated text fluency on WikiText Dataset~\cite{merity2016pointer}, and, \textit{Zero-shot Accuracy (Zero-shot Acc)} for next-token prediction evaluation on the mean of LAMBADA, HellaSwag, PIQA, and WinoGrande Datasets~\cite{eval-harness}; (4) \textbf{Attestation Overhead}: \textit{Latency overhead (\%)} is the additional time required to generate one token compared to pure REE inference, and \textit{Energy overhead (\%)} is the additional energy consumed to generate one token compared to pure REE inference.

\textit{\textbf{Baselines}.} \sys{} is the first attestation framework to protect on-device LLM's hardware IP. Nevertheless, we compare it with \textit{TEE-shield inference}~\cite{zhang2024no, liu2021trusted} that enforces model integrity and protects hardware IP by partitioning the LLM to fit into secure memory and executing segment by segment inside the TEE. 

Note that (i) we do not compare with DeepAttest~\cite{chen2019deepattest}, as it is designed for Intel SGX~\cite{costan2016intel} rather than Arm TrustZone. Besides, it encodes watermarks when training the DNNs from scratch, which cannot be easily scaled to LLMs~\cite{zhang2024emmark}; (ii) watermarking frameworks 
with intensive verification overheads are excluded because attestation requires a co-design of the watermarking algorithm and TEE system optimizations. Methods that perform full-model inference~\cite{li2023turning,li2024double} to check pre-encoded backdoors or require heavy extraction computation~\cite{zhang2024emmark} are therefore unsuitable for attestation.

\subsection{Results}
\paragraph{\textbf{Attestation reliability}} We encode watermarks into on-device LLMs from Llama, Qwen, and Phi families following Section~\ref{sec:watermark} and show the results in Table~\ref{tab:performance}. The attest pass rate (APR) reflects whether each layer's watermark can be correctly extracted. If any layer fails to achieve a 100\% layer-wise WER (i.e., APR lower than 100\%), the system flags the mismatch and aborts REE inference.
LLMs watermarked with \sys{} can reliably pass the attestation with 100\% WER and 100\% APR, whereas the execution of non-watermarked ones will be aborted due to low WER and 0\% APR.

\paragraph{\textbf{Watermarked LLM fidelity}}~\label{subsec:llm_perf}
Encoding watermarks onto LLMs will not degrade their quality and utility. We evaluate \sys's performance on text generation fluency using perplexity and next-token prediction capability using zero-shot accuracy.
As shown in Table~\ref{tab:performance}, for INT4 quantization, compared to non-watermarked models, \sys{} averagely degrades the Perplexity (PPL) by 0.62\% and Zero-shot Accuracy by 0.09\%. As for INT8 quantization, compared to non-watermarked models, \sys{} degrades the PPL and Zero-shot Accuracy by 0.07\% and 0.15\%, respectively. These degradations are minimal and do not produce noticeable differences when users interact with the watermarked models~\cite{lin2024awq}.

\begin{figure}[!ht]
    \centering
    \includegraphics[width=1\columnwidth]{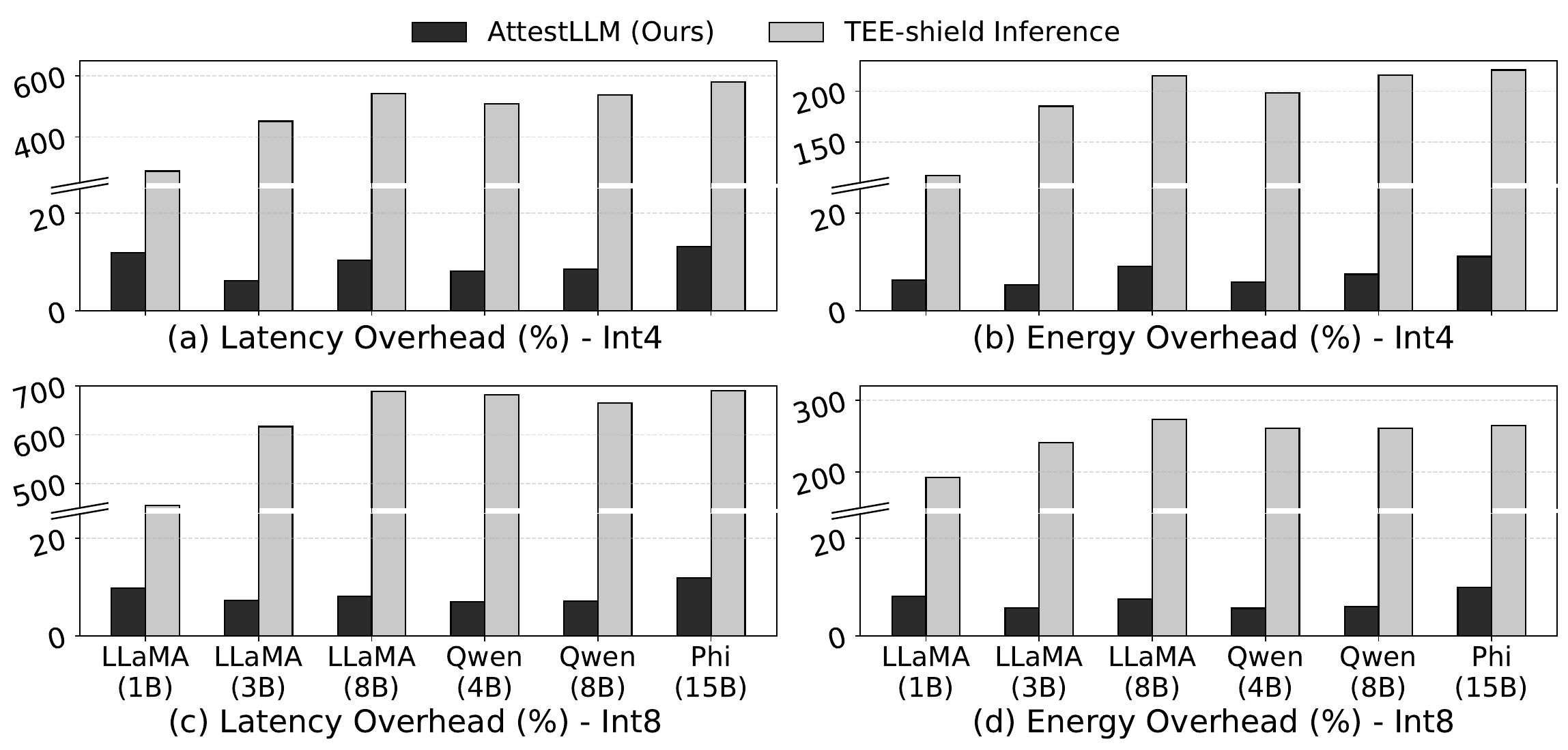}
    \vspace{-0.6cm}
    \caption{Attestation latency and energy overhead (\%) of LLMs quantized into INT4 and INT8. Lower is better.}

    \label{fig:overhead}
\end{figure}

\paragraph{\textbf{Attestation overhead}}~\label{subsec:overhead}
As shown in Figure~\ref{fig:overhead}, we report the attestation latency and energy overhead of \sys{} and TEE-shield inference compared to pure REE inference.  TEE-shield inference suffers from substantial overhead, with average latency overhead of 472.8\% for INT4 and 626.3\% for INT8, and average energy overhead of 87.8\% for INT4 and 247.2\% for INT8. In contrast, \sys{} imposes significantly lower costs over REE execution, with average overheads of 9.3\% (INT4) and 8.3\% (INT8) in terms of latency, and 7.3\% (INT4) and 7.0\% (INT8) for energy. As such, \sys{} achieves an average of $62\times$ lower latency overhead $30\times$ lower energy overhead compared to TEE-shield inference.


\paragraph{\textbf{Attestation strength}} 

The attestation is performed while ensuring sufficient attestation strength with a low evasion probability $P_{\text{evasion}}$.
For a conversational session producing 1000 tokens, with an attestation interval of $f{=}100$, \sys{} will conduct $r{=}10$  attestation rounds. Given that the model weights are stored in shuffled memory regions, we assume an adversary performs a brute-force search and tampers with 10\% of transformer layers in a Qwen3-8B model, which is 4 out of 36 layers. With Equation~\ref{eq:evasion}, the evasion probability per attestation is $6.1{\times}10^{-1}$, and over a single conversational session,  $P_{\text{evasion}}$ drops to $7.1{\times}10^{-3}$. It ensures \sys{} to confidently detect and abort the execution of unauthorized LLMs.  


\subsection{Robustness}~\label{subsec:attack}

We evaluate \sys's robustness by performing the following attacks: (i) watermark forgery attack, and (ii) model replacement attack; (iii) TEE system attack. We use Llama3-8B~\cite{grattafiori2024llama} model quantized to INT8 as the target model. We assume the adversary has prior knowledge of the watermarking algorithm, but do not have access to the encrypted trigger dataset and watermark keys. They can attempt a brute-force search to tamper with certain layers in the shuffled memory region, but locating and modifying all layers would be prohibitively expensive. 

\textit{\textbf{Watermark forgery attack}.} The adversary aims to host an unauthorized LLM but bypass the attestation by fooling the verification pipeline to decode the device-specific signatures. To achieve this, he/she needs to (i) know the watermark matrix $WM$ and watermark signature $B$, which are kept confidential and encrypted; (ii) brute-force search all of the LLM's parameters in shuffled and randomized memory, and replace them with the malicious copy. All of this information can be expensive to search and recover, which ensures \sys{} is resistant to watermark forgery attacks.

\textit{\textbf{Model replacement attack}.} 
In this setting, we consider two types of model replacement attacks (MPA): (i) partial model replacement attack; and (ii) full model replacement attack. In both cases, the adversary attempts to perform brute-force search and replaces model parameters with malicious copies. Since an exhaustive search in the shuffled memory region can be time-consuming, we consider partial model replacement as a more realistic threat, where $k=4$ out of $|M| = 32$ layers are tampered. The full model replacement attack is a stress test to evaluate \sys{}’s attestation reliability when all of the parameters are tampered. 
As in Table~\ref{tab:replacement_attack}, we use zeroth-order optimization to fine-tune INT8 quantized Llama3-8B~\cite{grattafiori2024llama} model on the ToxicChat~\cite{lin2023toxicchat} dataset. For partial MPA, we only update the parameters of four randomly selected layers; for full MPA, we update the entire model.  In the partial MPA case, four tampered layers fail attestation (i.e., layer-wise WER is not 100\%), resulting in an attest pass rate of 87.50\%. In such cases, the evasion probability after 10 rounds of attestation becomes $P_{\text{evasion}} = 3.5 \times 10^{-3}$, which ensures \sys{} to confidently abort the execution of tampered LLM.
In the full MPA case, all layers fail attestation because updating all of the model parameters alters their decoded watermarks.
These results demonstrate that \sys{} reliably detects both partial and full MPA, ensuring that unauthorized or tampered LLMs are blocked from execution.

\input{table/attack}

\paragraph{\textbf{TEE system attack}}  In TEE system attack, the adversary uses device resources to host an illegitimate LLM in REE. He/She aims to bypass attestation by (i) halting the interaction with TEE; (ii) dispatching the watermarked LLM solely for attestation. They can be mitigated by these system-level security efforts: (i) during platform bootstrapping, the hardware root-of-trust verifies the integrity of the bootloader, host kernel image, and device tree, preventing the injection of malicious software before the system starts; (ii) A secure timer inside the TEE tracks attestation intervals, which cannot be tampered with by the adversary. If the TEE does not receive attestation interactions within the expected time window, the hypervisor halts the application executions in REE, preventing the adversary from running a separate LLM without TEE involvement.; (iii) The attested layer selection is generated randomly and securely inside TEE, which prevents the attacker from manipulating layer checking and bypassing the attestation of tampered layers; (iv) The hypervisor monitors OS-level page allocations to ensure that the executed model matches the attested one. This prevents the adversary from running an illegitimate LLM while dispatching the watermarked model only for attestation.


\subsection{Ablation Studies and Analysis}~\label{ablation}

\textit{\textbf{Watermark before/after quantization}.} As in Table~\ref{tab:before_after_quantization}, when watermarking Llama3-1B~\cite{grattafiori2024llama} model quantized to INT4, post quantization WM insertion introduces more degradation because zeroth-order optimization must approximate gradients from a discrete weight distribution, which requires larger weight updates than gradient-based optimization. Pre-quantization WM insertion causes less quality degradation but suffers from reduced extraction accuracy, since quantization perturbs the activation outputs and weakens the watermark signal.
By combining both, \sys{} preserves both model quality and verification reliability.

\input{table/before_after_quantization}

\paragraph{\textbf{Attestation overhead with/without efficiency optimization}}
Without optimization, secure copy and watermark verification execute sequentially, leaving TEE threads idle and thereby increasing attestation overhead. As shown in Table~\ref{tab:optimization}, applying pipelined secure copy time overlapping and parallel LLM layer verification significantly improves efficiency. For Qwen3-4B quantized to INT8, \sys{}’s latency overhead decreases from 8.9\% to 6.9\%, and its energy overhead drops from 7.2\% to 5.7\%, demonstrating that these optimizations effectively reduce attestation cost.

\input{table/optimization}

\begin{figure}[!ht]
    \centering
    \includegraphics[width=0.9\columnwidth]{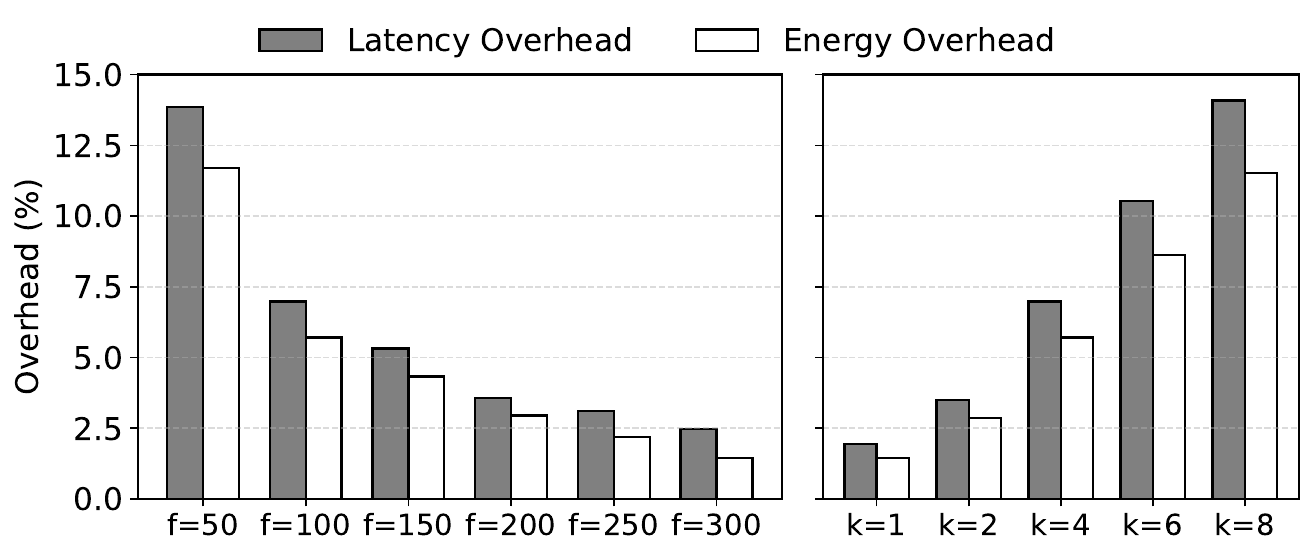}
    \caption{Attestation overhead with different ($f$, $k$) choices.
    } 
    \label{fig:sensitivity}
\end{figure}

\paragraph{\textbf{Attestation overhead with ($f$, $k$) choices}} $f$ and  $k$ are tunable hyperparameters that balance attestation strength and efficiency.  We show the quantitative results for Qwen3-4B quantized to INT8 in Fig.~\ref{fig:sensitivity}. A larger $f$ increases the attestation interval, reducing overhead but weakening security. Similarly, a smaller $k$ means fewer transformer layers are verified per round, lowering overhead but also reducing attestation strength.  Thus, we choose $f=100$ and $k=4$ for a reasonable trade-off between efficiency and security.


%% file: table/main_perf_table.tex
\begin{table*}[!htbp]
    \centering
   \resizebox{0.9\textwidth}{!}{%
    \begin{tabular}{c|c|ccc|cc|c|c|ccc|cc|c|c}
    \toprule
  \multirow{2}{*}{Method} &  Metrics&  \multicolumn{7}{c|}{INT4 Quantization} &  \multicolumn{7}{c}{INT8 Quantization }\\ \cline{2-16}
    \multirow{3}{*}{} &  \multirow{2}{*}{Model}  & \multicolumn{3}{c|}{Llama 3 } &  \multicolumn{2}{c|}{Qwen 3} & Phi-4  & \multirow{2}{*}{$\bar{\Delta}$}  & \multicolumn{3}{c|}{Llama 3 } &  \multicolumn{2}{c|}{Qwen 3} & Phi-4  & \multirow{2}{*}{$\bar{\Delta}$}\\
      &    & 1B & 3B & 8B & 4B & 8B & 15B & &1B & 3B & 8B & 4B & 8B & 15B &  \\\hline
  \multirow{4}{*}{non-WM} & PPL $\downarrow$ & 10.84 & 8.22 & 6.64 & 14.88& 9.98 & 6.78 & -& 9.88 & 7.88 & 6.36 & 13.82 & 9.63 & 6.75  & -\\ \cline{2-16}
  &   Zero-shot Acc $\uparrow$  & 59.59\% & 67.08\% & 71.38\% & 62.09\% & 66.16\% & 71.90\% &- & 60.85\% & 67.45\% & 72.11\% & 62.59\% & 67.27\% & 72.89\% & - \\\cline{2-16}
  &  WER $\uparrow$ & 55.50\% &  63.21\% & 63.68\% & 79.38\% & 79.41\% & 76.92\% & \cellcolor{gray!20}  & 71.42\% & 73.92\% & 67.71\% & 76.61\% & 81.33\% & 74.44\% & \cellcolor{gray!20} \\\cline{2-16}
  & APR $\uparrow$ & 0\% & 0\%  & 0\% & 0\% &0\%  & 0\%& \cellcolor{gray!20}  & 0\% & 0\% & 0\% & 0\% & 0\% &0\% & \cellcolor{gray!20} \\\hline
\multirow{3}{*}{\sys} 
    & PPL $\downarrow$& 10.92 &  8.24 & 6.66 & 15.07 & 9.96 & 6.85& 0.62\% & 9.92
& 7.89 & 6.37 & 13.72 & 9.62 & 6.76 & 0.07\%\\\cline{2-16} 
    &  Zero-shot Acc $\uparrow$ & 59.49\%  & 67.10\% & 71.18\% &61.97\% & 66.18\%
 & 71.89\%& 0.09\% & 61.31\% & 67.59\% & 72.04\% & 62.50\% & 67.66\% & 72.70\% & 0.15\%\\\cline{2-16}
    & WER $\uparrow$ & 100\% & 100\%  & 100\% & 100\% &100\%  & 100\%& \cellcolor{gray!20}  & 100\% & 100\% & 100\% & 100\% & 100\% &100\% &\cellcolor{gray!20}  \\\cline{2-16}
    & APR $\uparrow$ & 100\% & 100\%  & 100\% & 100\% &100\%  & 100\%&  \cellcolor{gray!20} & 100\% & 100\% & 100\% & 100\% & 100\% &100\% & \cellcolor{gray!20}  \\
    \bottomrule
    \end{tabular}}
    \caption{\sys's performance in watermarking on-device LLMs quantized to INT4 and INT8.  $\bar{\Delta}$ is the average LLM performance degradation (PPL for perplexity; Zero-shot Acc for zero-shot accuracy) compared with non-wm LLMs. }
    \label{tab:performance}
\end{table*}


%% file: table/attack.tex
\begin{table}[!htbp]
\centering
\small
\setlength{\abovecaptionskip}{2pt}
\vspace{-0.4cm}
\begin{tabular}{cccc}
\toprule
 \textbf{Attack} & \textbf{WM LLM (benign)} & \textbf{Partial MPA} & \textbf{Full MPA}  \\
\midrule
WER $\uparrow$ & 100\%  & 94.06\%  & 84.84\% \\
APR $\uparrow$ & 100\%  & 87.50\%  & 0\%\\
\bottomrule
\end{tabular}
\caption{Model replacement attack (MPA) performance.}
\label{tab:replacement_attack}
\vspace{-0.8cm}
\end{table}

%% file: table/before_after_quantization.tex
\begin{table}[!htbp]
\centering
\small
\vspace{-0.3cm}
\resizebox{0.9\columnwidth}{!}{%
\begin{tabular}{lcccc}
\toprule
\textbf{WM Insertion Stage} & \textbf{PPL} $\downarrow$ & \textbf{Zero-shot Acc} $\uparrow$ & \textbf{WER} $\uparrow$ & \textbf{APR} $\uparrow$\\
\midrule
Before Quantization &  10.26 & 59.58\% & 97.19\% & 87.50\%\\
After Quantization & 10.84 & 58.51\% & 100\%& 100\%\\
\sys{}  & 9.91 & 60.90\% & 100\%& 100\%\\
\bottomrule
\end{tabular}}
\caption{Watermark insertion before/after quantization.}
\label{tab:before_after_quantization}
\vspace{-0.9cm}
\end{table}

%% file: table/optimization.tex
\begin{table}[!htbp]
\centering
\small
\vspace{-0.3cm}
\begin{tabular}{ccc}
\toprule
\textbf{Efficiency Optimization} & \textbf{Latency} $\downarrow$ & \textbf{Energy}  $\downarrow$ \\
\midrule
no Optimization & 8.9\% &  7.2\% \\
with Optimization  & 6.9\% & 5.7\% \\ 
\bottomrule
\end{tabular}
\caption{Attestation overhead with/without optimizations.}
\vspace{-0.8cm}
\label{tab:optimization}
\end{table}